# Stock mechanics: energy conservation theory and the fundamental line of DJIA

Çağlar Tuncay
Department of Physics, Middle East Technical University
06531 Ankara, Turkey
caglart@metu.edu.tr

**Abstract**

DJIA is tested whether it can be described as a mechanical system conserving total energy with K ($=\frac{1}{2}v^2$) + U, where U is calculated as the negative of work done by force obtained in terms of the second derivative of price, assuming unit mass.

## 1. Introduction

Market indices as well as share prices, during their time excursions pass through several states; oscillations, rises and falls, crashes (crises), etc. During that time, some quantities may remain as conserved, such as total energy. As a result, time series may be predicted epoch by epoch, in terms of some simple analytical functions resulting from energy conservation, once the potential energy of the current epoch is known. In this work we will exemplify an application of energy conservation theory (e-ct) on DJIA in the "$0^{th}$-order approximation". Some possible extensions of energy conservation theory will be pronounced in the final section.

## 2. Energy conservation theory

The whole history of DJIA, utilizing daily real data [1] can be decomposed into five main epochs, each with a different length of duration. Refering to Fig. 1. the first epoch is the transitional period lasted about 3565 days from the establishment of the index on. The main characteristics of this epoch is (almost) triangular squeezing of prices with the climax meeting the beginning of the Sept.1929 crash ("Black Monday") at a value of 381.17, which ended after about 700 days at a value of 41.22, (Fig. 2.). Then a sharp rise occured till the value of about 200. Afterwards some decrease came reaching its deepest value at 92.92. These two successive local minima, i.e., 41.22 and 92.92 are very important figures in the history of DJIA since, the line passing through them defines the lower rising edge of the pronounced triangle and moreover the rightward extension of this line plays the role of carrying (supporting) prices (i.e. the line passing through minima) for many decades, till the value increases to nearly 1000. (See, the longest straight line in Fig. 1.)

The pronounced line is in fact worth to be considered as the fundamental line of DJIA because of several reasons. First of all, it can be taken as the $0^{th}$-order approximation for the index, since the corresponding time series is observed to evolve about (below and above of) it. Secondly the fundamental line, has played the role of supporting line till the date of 16.June.1969 ($10{,}172^{th}$ day). The day after, i.e. at 17.June.1969 the index has broken it down (after many previous hits) and later has never crossed it. Starting in Apr.1999 DJIA has approached it from below many times, (at index values between 11,000 and 11,700) till the end of the same year, but has ceased to override it, and the last log-periodic crash has followed. (For detailed studies of this crash, in a power law formalism, see [2-6].) Slope ($\beta$) of the present log-linear increase of the fundamental line is calculated to be $\beta = \ln(92.92/41.22)/(3390-946)=0.000333\text{day}^{-1}$. Therefore, the equation of line can be written as

$$\chi(t)=\chi_0 \exp(0.000333\, t) \quad , \qquad (1)$$

the constant $\chi_0$ takes care of t=0 value, i.e., the initial price level. $\chi_0$ may very well be taken as 41.22 or 92.92 and the date of the corresponding day will be the origin for time.

We have the price velocity and acceleration as $v = d\chi/dt = \beta\chi$, and $a = dv/dt = d^2\chi/dt^2 = \beta^2\chi$, respectively, and both increase exponentially in time. Assuming m=1[7], one may write an explicit function for the $0^{th}$-order force

$$F^{\textit{zeroth-order}} = a = (0.000333)^2 \chi_0 \exp(0.000333\, t) \quad . \qquad (2)$$

Once the force is known, one may calculate the work done by this force, along the path of Eq. (1) that DJIA follows within the $0^{th}$-order approximation: $\Delta W=\int F dx=\int \beta^2\chi\, d\chi = \frac{1}{2}\beta^2\chi^2$. The present indefinite integral may be converted into a definitine one by substituting the initial $\chi$-value for t=0, and the general term for any t; therefore $\Delta W = \frac{1}{2}\beta^2(\chi^2 - \chi_0^2)$, which is nothing but $-\Delta U$, by definition. Therefore,

$$\Delta U = -\tfrac{1}{2}\beta^2 (\chi^2 - \chi_0^2) \quad . \qquad (3)$$

Since, $v = \beta\chi$; then one may calculate the change in kinetic energy in the same journey of DJIA, as $\Delta K= \frac{1}{2} (v^2 - v_0^2) = \frac{1}{2}\beta^2(\chi^2 - \chi_0^2)$. So,

$$\Delta U + \Delta K = 0 = \Delta(U + K) \quad . \qquad (4)$$

Please note that, Eq. (4) is a result of the exponential growth (Eq. (1)) of DJIA. There might be found some other quantities as being conserved during the same time excursion of DJIA, and within world market. With respect to this possibility, it is better to have a separation between any conservation theory and the current energy conservation theory (e-ct).

Energy conservation (Eq. (4)) implies absence of any frictional force or damping etc. on DJIA, along the way under consideration. Yet, DJIA is observed to deviate from its usual exponentially increasing path, and spend about twenty years (1962-1982) below the index value of 1000. During that time, it bounced up and down many times between the levels of 750 and 1000 as displayed in Fig. 3. Inspiring also from the quadratic form of potential energy in Eq. (3), we may approximate this oscillatory mood by a sinusoidal (harmonic) form. Utilizing the observational data, the following expression may be proposed to represent the third epoch as

$$\chi(t)^{epoch\ 3} = 875 + 125\sin((2\pi/750)(t - t_0) - \pi/2) \ , \qquad (5)$$

where, $t = t_0 = 0$ may be taken at the date of 11.Apr.1963, at an index value of 750, and the angular frequency (w) comes out as $2\pi/750 \cong 7{,}02\mathrm{x}10^{-5}$ $\mathrm{day}^{-1}$.

It can be easily shown that, $\Delta U= \frac{1}{2}(2\pi/750)^2(\chi^2 - \chi_0^2)$. Furthermore, $v=\beta\chi$, and $a=\beta^2\chi=F$ relations still hold with $\beta^2 = -w^2$. Total energy $E_T = U + K$ of an harmonic oscillator is known be $\frac{1}{2}(2\pi/T)^2A^2$, where T and A stand for period and the amplitude, respectively. Furthermore $\Delta E = \Delta U + \Delta K = 0$. Subsituation of the observed data yields $E_T=0.55$ $(\mathrm{value/day})^2$.

As a closer investigation on Fig. 3. delivers the fact that, Eq. (5) does not fit the real data everywhere. For example, the minima of zigzags fall down twice below the 750 level. And about the middle, real maximum occurs when the sinusoidal has its minimum, and vice versa right after it, i.e., real minimum occurs when the sinusoidal is at its maximum. Yet, this deviation may be expressed within the present context of energy conservation; as the real period exceeds that of the sinusoidal (750 day), the local amplitude increases, and as a result A/T remains the same, conserving totat energy $E_T=\frac{1}{2}(2\pi A/T)^2$.

At the end of the year of 1982, DJIA overcomes the "psychological" resistance of 1000, and exponentially increases again, with a steeper slope of $\beta=0.000500$ $\mathrm{day}^{-1}$, till it hits the fundamental line from below at Jan.1999. During that time, total energy (with $U=-\frac{1}{2}(0.0005)^2\chi^2$) remains constant. Afterwards, the 2000 crash comes.

As it can be seen in Fig. 1., nowadays DJIA is at maximum distance from the fundamental line, and it seems as if that the distance will continue to increase in close future. The current epoch, may be a new transitional one. Yet, one may represent it by a gravitational potential energy [7].

## Conclusion

The three epochs out of five, are shown to involve a quadratic potential energy in the $0^{th}$-order approximation. Conservation of total energy is proved in terms of real data $\chi$ only. Force equations are obtained for the three median epochs. For finer details of the excursion of DJIA, some more potential terms may be added, solutions of which can be obtained via perturbational techniques as $1^{st}$- and higher-order approximations to DJIA. The existing U and K scales by some arbitrary $\lambda^2$, whereas $\chi$, v, and a scales by $\lambda$, proving a power law.

Furthermore, once the potential and total energies are obtained (e.g. for exponential behavior U and $E_T$ will be $-\frac{1}{2}\beta^2\chi^2$ and zero, respectively; for oscillations $U=+\frac{1}{2}\beta^2\chi^2$, and $E_T=2(\pi A/T)^2$, and for azimuthal rises and falls $U=h_1\chi$, and $E_T= h_1\chi_0$) one may try Lagrangian and even wave mechanical solutions. Just to exemplify, for exponential growth we will have $\{(-\frac{1}{2}\hbar^2)(d^2/d\chi^2) - \frac{1}{2}\beta^2\chi^2\}\psi = 0$, for oscillations we will have $\{(-\frac{1}{2}\hbar^2)(d^2/d\chi^2) + \frac{1}{2}\beta^2\chi^2\}\psi = 2(\pi A/T)^2\psi$, and for azimuthal rises and falls we will have $\{(-\frac{1}{2}\hbar^2)(d^2/d\chi^2) + h_1\chi\}\psi = h_1\chi_0\psi$. Furthermore, application of boundary conditions may yield some kind of quantization in some quantities.

## Acknowledgement

The author is thankful to Dietrich Stauffer for his inspirations and friendly discussions and corrections.

## Figure captions

Fig. 1. DJIA, with a logarithmic price axis. Five fundamental epochs composing its history can be selected. The first “transitional” epoch is characterized by its triangular squeezing the price about the first “psychological resistance” of the value of 100. In the second and the fourth epochs log-linear (i.e., exponential) ascending of the price can be distinguished (arrows 1 and 2). In the third epoch, just below the second “psychological resistance” of the value of 1,000 oscillations are dominant. In the fifth, and the final epoch, the evolution of formation (about the third “psychological resistance” of 10,000 index) is not completed yet.

Fig. 2. The first transitional epoch in DJIA. Notice the supporting line, rightward extention of which displays that it is fundamental for DJIA.

Fig. 3. The third, (oscillatory) epoch of DJIA. When the time difference between any successive hits to the supporting level of 750 (i.e., the period T) increases, the amplitude (A) also increases. As a result A/T ratio, and $E_T$ remains the same.

# Figures

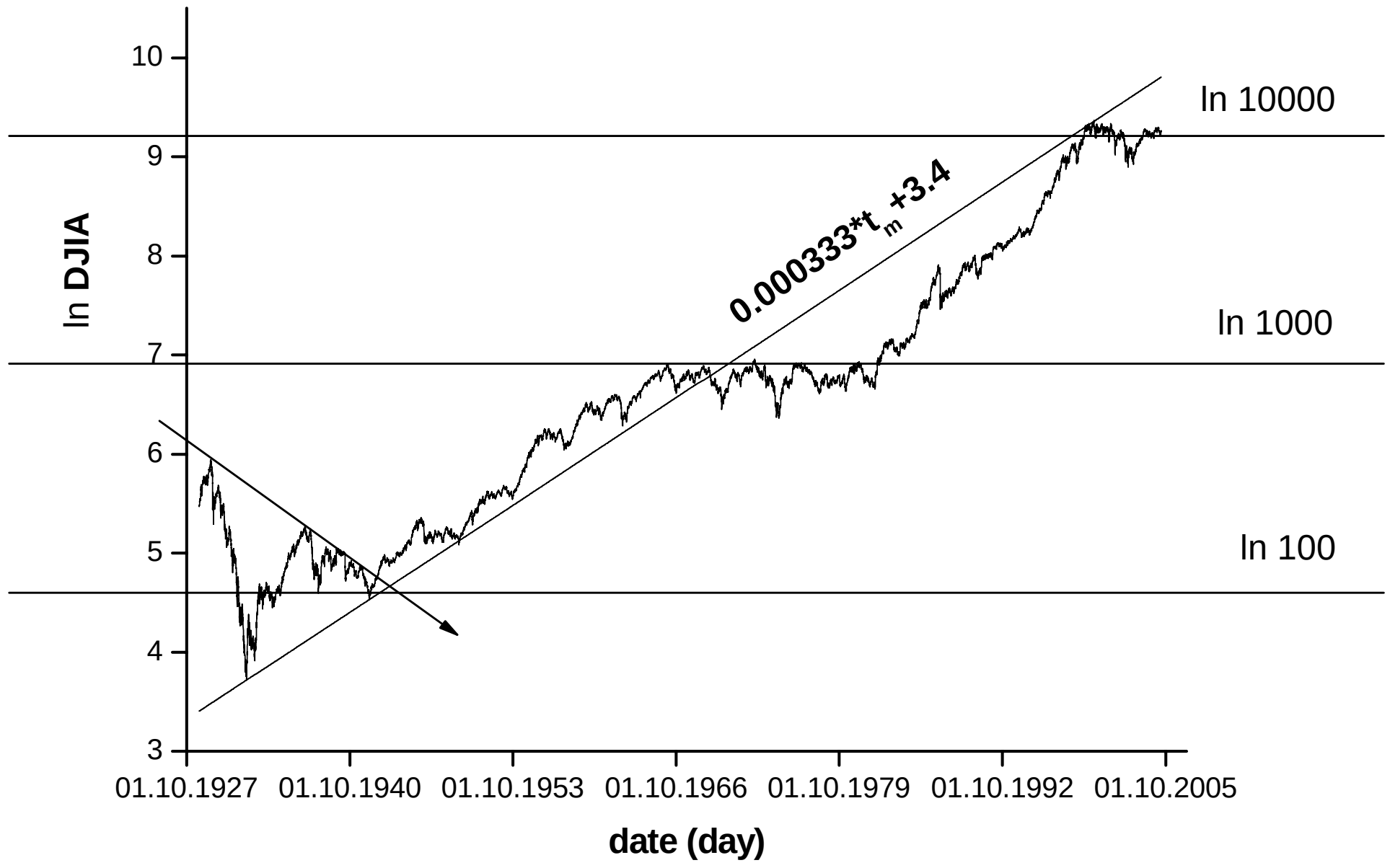

**Figure 1.**

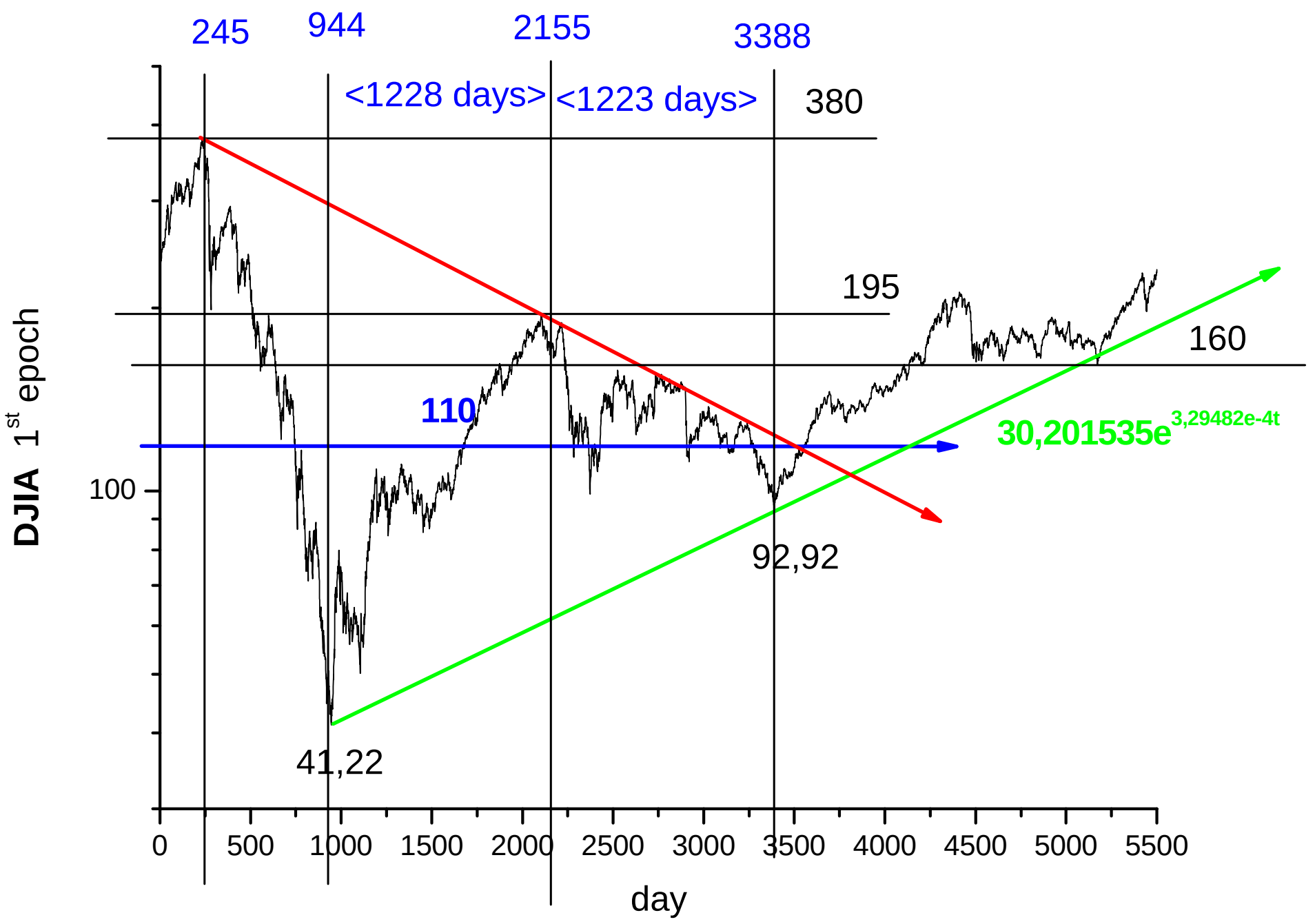

**Fig. 2.**

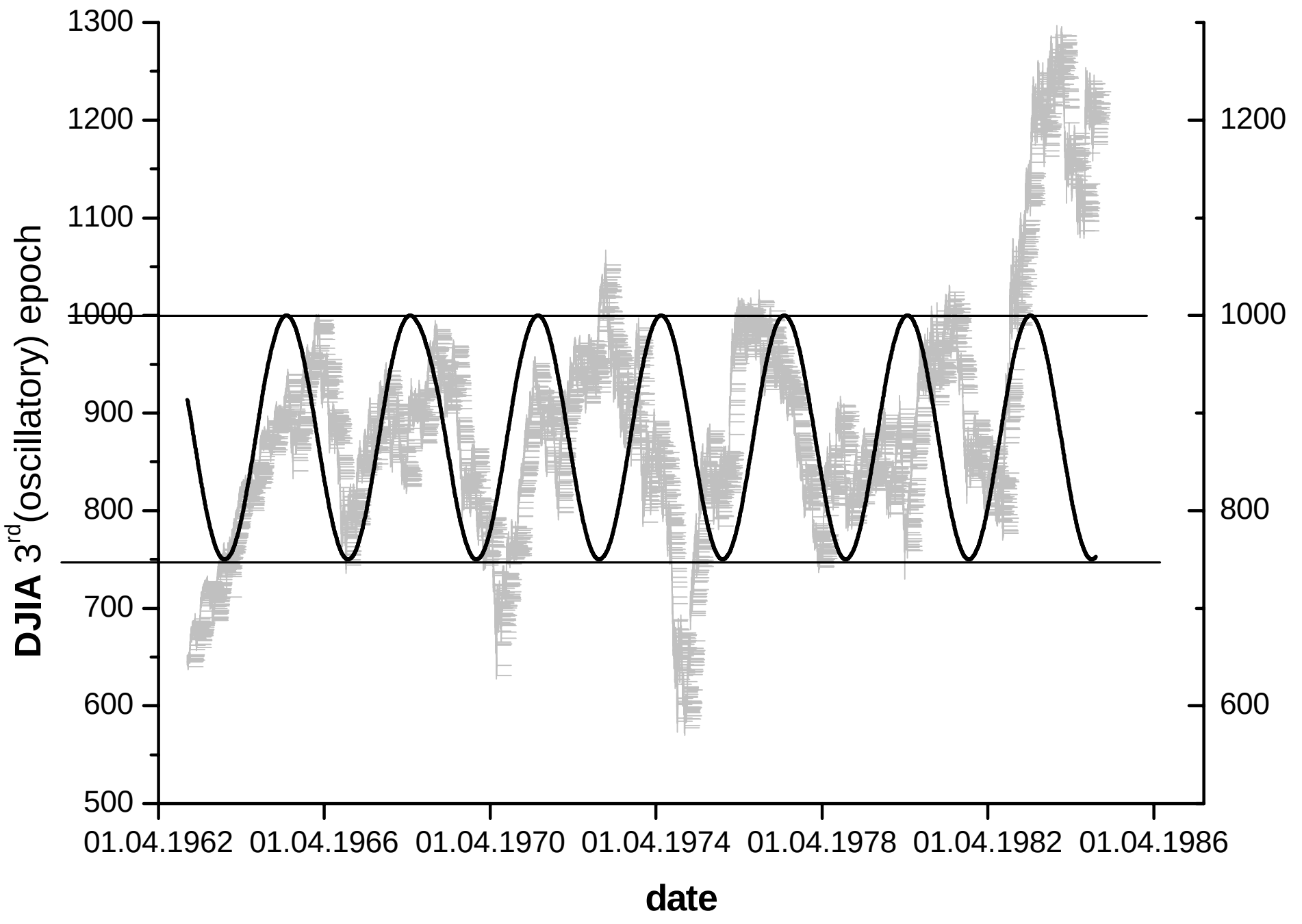

**Fig. 3.**